\documentclass{aa}

\usepackage{graphicx}
\usepackage{natbib}
\bibpunct{(}{)}{;}{a}{}{,}

\begin{document}
\title{The gas temperature in flaring disks around pre-main sequence stars}

\author{B. Jonkheid\inst{1}\and F.G.A. Faas\inst{1}\and G.-J. van Zadelhoff\inst{1,2}\and E.F. van Dishoeck\inst{1}}

\authorrunning{Jonkheid et al.}
\titlerunning{The gas temperature in flaring disks}

\offprints{B. Jonkheid, \email{jonkheid@strw.leidenuniv.nl}}

\institute{
Sterrewacht Leiden, P.O. Box 9513, 2300 RA Leiden, the Netherlands
\and Royal Dutch Meteorological Institute, P. O. Box 201, 3730 AE De Bilt, the Netherlands}

\date{Received 5 April 2004  / Accepted 23 August 2004}

\abstract{A model is presented which calculates the gas temperature and 
chemistry in the surface layers of flaring circumstellar disks using a 
code developed for photon-dominated regions. Special
attention is given to the influence of dust settling.  It is found
that the gas temperature exceeds the dust temperature by up to several
hundreds of Kelvins in the part of the disk that is optically thin to
ultraviolet radiation, indicating that the common assumption that 
$T_{\rm gas}=T_{\rm dust}$ is not valid throughout the disk. In the 
optically thick part, gas and dust are
strongly coupled and the gas temperature equals the dust
temperature. Dust settling has little effect on the chemistry in the
disk, but increases the amount of hot gas deeper in the disk. The 
effects of
the higher gas temperature on several emission lines arising in the
surface layer are examined.  The higher gas temperatures increase the
intensities of molecular and fine-structure lines by up to an order of
magnitude, and can also have an important effect on the line shapes.

}

\maketitle

\section{Introduction}
Circumstellar disks play a crucial role in both star and planet
formation.  After protostars have formed, part of the remnant material
from the parent cloud core can continue to accrete by means of viscous
processes in the disk \citep{shu87}. Disks are also the sites of
planet formation, either through coagulation and accretion of dust grains 
or through gravitational instabilities in the disk \citep{liss93,boss00}.
To obtain detailed information on these processes, both the dust and the
gas component of the disks have to be studied. Over the last decades,
there have been numerous models of the dust emission from disks, with
most of the recent work focussed on flaring structures in which
the disk surface intercepts a significant part of the stellar
radiation out to large distances and is heated to much higher
temperatures than the midplane 
\citep[e.g.][]{kenhar87,chigol97,dale98,dale99,bell97,dull02}. These 
models are appropriate to the later stages of the disk evolution when 
the mass accretion rate has dropped and the remnant envelope has dispersed, 
so that heating by stellar radiation rather than viscous energy release 
dominates. Such models have been shown to reproduce well the observed 
spectral energy distributions at mid- and far-infrared wavelengths for a 
large variety of disks around low- and intermediate-mass pre-main sequence 
stars.

Observations of the gas in disks started with millimeter
interferometry data of the lowest transitions of the CO molecule
\citep[e.g.][]{koesar95,dutr96,mansar97,dart03}, but now also include 
submillimeter single-dish \citep[e.g.][]{kast97,thi01,zade01} and infrared 
\citep[e.g][]{najita03,brit03} data on higher excitation CO lines. 
Evidence for the
presence of warm gas in disks also comes from near- and mid-infrared 
and ultraviolet observations of the H$_2$ molecule 
\citep[e.g.][]{herc02,bary03,thi01}. Although emission from
the hottest gas probed at near-infrared and ultraviolet wavelengths is
thought to come primarily from a region within a few AU of the young
stars, the longer wavelength data trace gas at larger distances from
the star, $>$50 AU.  Molecules other than CO are now also detected at
(sub)millimeter wavelengths, including CN, HCN, HCO$^+$, CS and
H$_2$CO \citep[e.g.][]{dutr97,kast97,qi03,thi04}.

The emission from all of these gas tracers is determined both by their
chemistry and excitation, where the latter depends on the temperature
and density structure of the disk.  The chemistry in flaring disks has
been studied intensely in recent years by various groups
\citep[e.g.][]{aikume97,willan00,mark02,zade03}, whereas the density 
structure is constrained from vertical
hydrostatic equilibrium models of the dust disk assuming a contant
gas/dust ratio \citep{dale99,dull02}. In all of these models, the gas
temperature is assumed to be equal to the dust temperature. That this
assumption does not always hold was shown by \citet{kamzad01}, who
calculated the gas temperature in tenuous disks around Vega-like
stars. These disks have very low disk masses, of order a few $M_\oplus$, 
and are optically thin to ultraviolet (UV) radiation
throughout. It was shown that their gas temperature is generally very
different from the dust temperature, and that this higher temperature
significantly affects the intensity of gaseous emission lines, in the
most extreme cases by an order of magnitude or more \citep{kamp03}.

In this work, the gas temperature in the more massive disks (up to 0.1
$M_{\rm Sun}$) around T-Tauri stars is examined, which are optically
thick to UV radiation. Special attention is given to the effects of
dust settling and the influence of explicit gas temperature
calculations on the properties of observable emission lines. 

The model used in this paper is described in Section 2. The resulting 
temperature, chemistry and emission lines are shown in Section 3. In 
Section 4 the limitations of our calculations are discussed, and in 
Section 5 our conclusions are presented. The details of the heating and 
cooling rates are in Appendices A and B, respectively.

\section{Calculating the gas temperature}
\subsection{The PDR model}

\begin{figure}
\resizebox{\hsize}{!}{\includegraphics[angle=0]{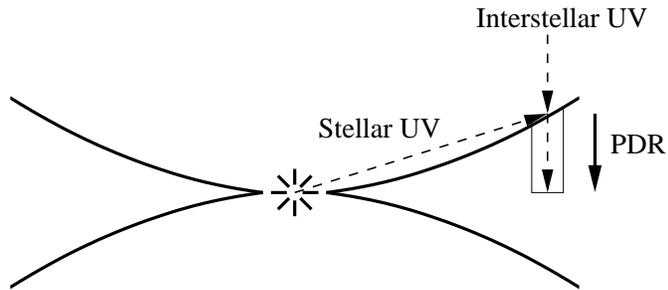}}
\caption{The 1+1-D model. The disk is divided into annuli, each of 
which is treated as a 1-dimensional PDR problem for the chemistry and the 
cooling rates.}
\label{1+1d}
\end{figure}

Because of the symmetry inherent in the disk shape, it is usually
assumed that disks have cylindrical symmetry, resulting in a
2-dimensional (2-D) structure. Because of the computational problems in
solving the chemistry (particularly the ${\rm H_2}$ and CO self shielding) and 
cooling rates in a 2-D formalism, however, these
are calculated by dividing the 2-D structure into a series of 1-D structures 
(see Figure \ref{1+1d}). The disk is divided
into 15 annuli between 50 and 400 AU. These 1-D structures resemble the 
photon-dominated regions (PDRs) found at the edges of molecular clouds 
\citep[for a review, see][]{holtie97}. A full 2-D formalism is used to 
calculate the dominant heating rates due to the photoelectric effect on 
Polycyclic Aromatic Hydrocarbons (PAHs) and small grains.

The 1-D PDR model described by \citet{bladis87}, \citet{disbla88} and
\citet{jans95} is used in this work. This code consists of two parts:
the first part calculates the photodissociation and excitation of
${\rm H_2}$ in full detail, taking all relevant H$_2$ levels and lines
into account. It includes a small chemical network containing the
reactions relevant to the formation and destruction of ${\rm H_2}$ to
compute the H/H$_2$ transition. The main output from this code is the
H$_2$ photodissociation rate as a function of depth and the fraction
H$_2^*$ of molecular hydrogen in vibrationally-excited states.  The
second part of the program includes a detailed treatment of the
CO photodissociation process, a larger chemical network to determine
the abundances of all species, and an explicit calculation of
all heating and cooling processes to determine the gas temperature.
The latter calculation is done iteratively with the chemistry, since
the cooling rates depend directly on the abundances of O, C$^+$, C and
CO. Each PDR consists of up to 130 vertical depth steps, with a variable
step size adjusted to finely sample the important H$\to$H$_2$ and
C$^+$$\to$C$\to$CO transitions.

To simulate a circumstellar disk, some modifications had to be made in
the PDR code. Because of the higher densities, thermal coupling between
gas and dust had to be included. While this is not an important factor
in most PDRs associated with molecular clouds, it dominates the
thermal balance in the dense regions near the midplane of a disk. A
variable density also had to be included to account for the increasing
density towards the midplane of the disk. As input to the code, the
density structure as well as the dust temperature computed by
\citet{dale99} are used. In this model, the structure is calculated
assuming a static disk in vertical hydrostatic equilibrium. The disk
is considered to be geometrically thin, so radial energy transport can
be neglected. The model further assumes a constant mass accretion rate
throughout the disk (a value of $\dot{M}=10^{-8}\,M_\odot\,{\rm
yr^{-1}}$ is used), and turbulent viscosity given by the $\alpha$
prescription ($\alpha=0.01$). In the outer disk studied here, the
contribution of accretion to the heating is negligible. The central
star is assumed to have a mass of $0.5\,M_\odot$, a radius of
$2\,R_\odot$ and a temperature of 4000 K. The disk mass is
$0.07\,M_\odot$ and extends to $\sim$400 AU. This same model has been
used by \citet{aikawa02} and \citet{zade03} to
study the chemistry of disks such as that toward LkCa 15 and TW
Hya. As shown in those studies, the results do not depend strongly on
the precise model parameters.

Further input to the PDR code is the strength of the UV field. The UV field 
is based on the spectrum for TW Hya obtained by 
\citet{costa00} \citep[stellar spectrum B in][]{zade03}: a 4000 K 
black body spectrum, plus a free-free and
free-bound contribution at $3\times 10^4\ {\rm K}$ and a 7900 K
contribution covering 5\% of the central star. The intensities of the
resulting UV radiation incident on the disk surface at each annulus
(with both a stellar and an interstellar component) are calculated
using the 2-D Monte Carlo code of \citet{zade03}. The resulting UV
intensities are then converted into a scaling factor $I_{\rm UV}$
with respect to the standard interstellar radiation field as given by
\citet{draine78}. These scaling factors $I_{\rm UV}$ form the input to 
calculate the chemistry for each of the 15 vertical PDRs and range 
from $\sim 1000$ at the first slice at 63 AU
to $\sim 100$ at the last slice at 373 AU. As shown by \citet{zade03}, 
the difference in the chemistry calculated with spectrum B and a scaled 
Draine radiation field (spectrum A) is negligible. It is important to note 
that both radiation fields have ample photons in the 912--1100 \AA \ regime
which are capable of photodissociating H$_2$ and CO and photoionizing
C. 

The UV radiation also controls the heating of the gas through the 
photoelectric effect on grains and PAHs. The rates of these two processes are 
calculated using the 
2-D radiation field at each radius and height in the disk computed by 
\citet{zade03} including absorption and scattering.

\subsection{Chemistry}

To calculate the chemistry, the network of \citet{jans95} is used. It
contains 215 species consisting of 26 elements (including isotopes)
with 1549 reactions between them.  The adopted abundances of the most
important elements are shown in Table \ref{abuntab}; the carbon and
oxygen abundances are similar to those used by \citet{aikawa02}.
The chemistry has been checked against updated UMIST databases, but only 
minor 
differences have been found for the species observed in PDRs. Since the 
temperature depends primarily on the abundances of the main coolants, the 
precise chemistry does not matter as long as the C$^+$$\to$C$\to$CO 
transition is well described. The adopted cosmic ray ionization rate is 
$\zeta=5\times 10^{-17}$ s$^{-1}$.

Since the chemical timescales in the PDR layer are short (at most a
few thousand yr) compared to the lifetime of the disk, the chemistry is 
solved in steady-state. Only gas-phase reactions are considered; no 
freeze-out onto grains is included. Accordingly, the PDR calculations 
are stopped when the dust temperature falls below 20~K. Since this is 
also the regime where the gas and dust are thermally coupled 
(see \S 3.1), no explicit calculation of the gas temperature is needed.

\begin{figure}
\resizebox{\hsize}{!}{\includegraphics[angle=0]{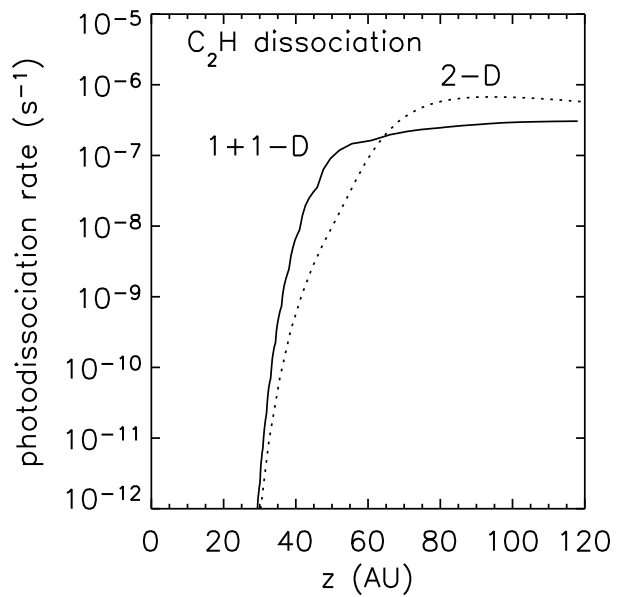}}
\caption{The photodissociation rate of ${\rm C_2H}$ at a radius of 105 AU. The solid line gives the rate found in this work, the dotted line the rate by \citet{zade03}.}
\label{disrates}
\end{figure}

It should be noted that only the vertical attenuation is considered
here in the treatment of the photorates used in the chemistry, in 
contrast to the work by
\citet{aikawa02} who included attenuation along the line of sight to the 
star. This means that the photorates are generally larger 
deeper into the disk. This is illustrated in Figure \ref{disrates}, where
the dissociation rate of ${\rm C_2H}$ is shown as 
functions of height. Comparison with the rate found by \citet{zade03}
shows that the dissociating radiation penetrates deeper
(to a height of $\sim 40$ AU opposed to $\sim 60$ AU) into the disk. Given 
the overall uncertainties in the radiation field and disk structure these 
effects are not very significant. Another difference of our model with those 
of \citet{aikawa02} and \citet{zade03} is the full treatment of CO
photodissociation as given by \citet{disbla88}, including self shielding and 
the  mutual shielding by ${\rm H_2}$.

\begin{table}
\centering
\caption[]{Adopted gas-phase elemental abundances with respect to hydrogen.}
\label{abuntab}
\begin{tabular}{l l}
\hline
Element & abundance\\
\hline
D & $1.5\times 10^{-5}$\\
He & $7.5\times 10^{-2}$\\
C & $7.9\times 10^{-5}$\\
N & $2.6\times 10^{-5}$\\
O & $1.8\times 10^{-4}$\\
S & $1.7\times 10^{-6}$\\
PAH & $1.0\times 10^{-7}$\\
\hline
\end{tabular}
\end{table}

\subsection{Thermal balance}

The gas temperature in the disk is calculated by solving the balance
between the total heating rate $\Gamma$ and the total cooling rate
$\Lambda$. The equilibrium temperature is determined using Brent's
method \citep{numrec}. Heating rates included in the code are due to
the following processes: photoelectric effect on PAHs and large
grains, cosmic ray ionization of H and ${\rm H_2}$, C photoionization,
${\rm H_2}$ formation, ${\rm H_2}$ dissociation, ${\rm H_2}$ pumping
by UV photons followed by collisional de-excitation, pumping of [O I]
by infrared photons followed by collisional de-excitation, exothermic
chemical reactions, and collisions with dust grains when $T_{\rm
dust}>T_{\rm gas}$. The gas is cooled by line emission of ${\rm
C^+}$, C, O and CO, and by collisions with dust grains when $T_{\rm
dust}<T_{\rm gas}$. A detailed description of each of these processes
is given in Appendix A. The thermal structure found in our PDR models
has been checked extensively against that computed in other PDR codes.

There are two important assumptions in our calculation of the thermal
balance which may not necessarily be valid for disks compared with the
traditional PDRs associated with molecular clouds. The first is the
use of a 1-D escape probability method for the cooling lines. This
limitation is further discussed in \S 4.1. The second is the
assumption that the photoelectric heating efficiency for interstellar
grains also holds for grains in disks. As shown by \citet{kamzad01},
this efficiency is greatly reduced if the grains have grown from the
typical interstellar size of 0.1 $\mu$m to sizes of a few $\mu$m.
While there is good observational evidence for grain growth for older,
tenuous disks, the younger disks studied here are usually assumed to
have a size distribution which includes the smaller grains.

\subsection{Dust settling}

In turbulent disks, the gas and dust are generally
well-mixed. \citet{steval96} have shown that the velocities of dust
grains with sizes $< 0.1$ cm are coupled to the gas. When the disk
becomes quiescent, models predict that dust particles are no longer
supported by the gas and will start to settle towards the midplane of
the disk \citep{weiden97}. During the settling process, dust particles
will sweep up and coagulate other dust particles, thus leading to
grain growth. Since larger particles have much shorter settling times
than small particles, coagulation accelerates the settling process.

Observational evidence for dust growth and settling in disks includes
the near-IR morphology of the edge-on disks. For example, for the
114-526 disk in Orion, \citet{throop01} and \citet{shupin03} show that
the large dust grains in this disk are concentrated in the
midplane. The SEDs of some of the T-Tauri stars examined by
\citet{miynak95} also indicate that dust settling takes place,
as well as the statistics of observations of edge-on disks by
\citet{dale99}.

In our model, dust settling is simulated by varying the gas/dust mass
ratio.  In this paper, both models with a constant $m_{\rm gas}/m_{\rm
dust}$ (``well-mixed'') and a variable value of $m_{\rm gas}/m_{\rm
dust}$ (``settled'') are presented. In the well-mixed case, the
gas/dust mass ratio is taken to be 100 throughout.  For the settled
model, the value of 100 is kept near the midplane of the disk, while
a value of $10^4$ is used in the surface regions.  The
precise value adopted at the surface is not important, as long as it
is high enough that photoelectric heating is no longer significant in
this layer.  The ratio is varied linearly with depth in a narrow
transition region. The boundaries of the transition region are
defined as $z_6\pm h/10$, where $z_6$ is the height where the settling
time of dust particles is $10^6$ years (this method thus simulates a
disk that is $10^6$ years old) and $h$ is the height of the D'Alessio 
disk, defined as the height where the gas pressure is equal to 
$7.2\times 10^{5}\,{\rm K\, cm^{-3}}$.

\begin{figure}
\resizebox{\hsize}{!}{\includegraphics[angle=0]{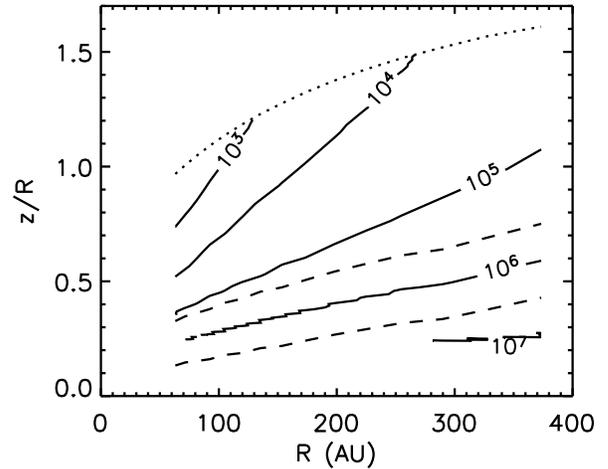}}
\caption{Settling times in years for 0.1 $\mu$m-sized dust particles. The 
dashed lines give the boundaries between which the $m_{\rm gas}/m_{\rm dust}$ 
ratio is interpolated from a value of $10^4$ to $10^2$.}
\label{settle}
\end{figure}

The settling time $t_{\rm settle}$ is determined as follows:
$$t_{\rm settle}=\frac{z}{v_{\rm settle}}$$ 
where $z$ is the height above the midplane. The settling velocity
$v_{\rm settle}$ is determined by the balance between the vertical
component of the gravitational force on the dust grains and the drag
force the grain experiences as it moves through the gas. For the drag
force the subsonic limit of the formulae by \citet{berruy91} is used,
resulting in a settling velocity of
$$v_{\rm
settle}=\frac{\sqrt{\pi}\,G\,M\,z\,a\,\rho}{2\,r^3\,\mu\,n_{\rm
H}\,v_{\rm thermal}}$$ 
with $G$ the gravitational constant, $M$ the
stellar mass, $a$ the grain radius (a typical value of 0.1 $\mu$m is
used), $\rho$ the density of the grain material (2.7 g cm$^{-3}$), 
$r=\sqrt{z^2 + R^2}$
the distance from the star, $\mu$ the mean molecular weight (2.3
proton masses) and $v_{\rm thermal}$ the thermal velocity of the gas
particles ($v_{\rm thermal}=\sqrt{2\,k\,T/\mu}$). The settling times
found by this method are displayed in Figure~\ref{settle}.

\begin{figure*}[!tp]
\centering
\includegraphics[width=17cm]{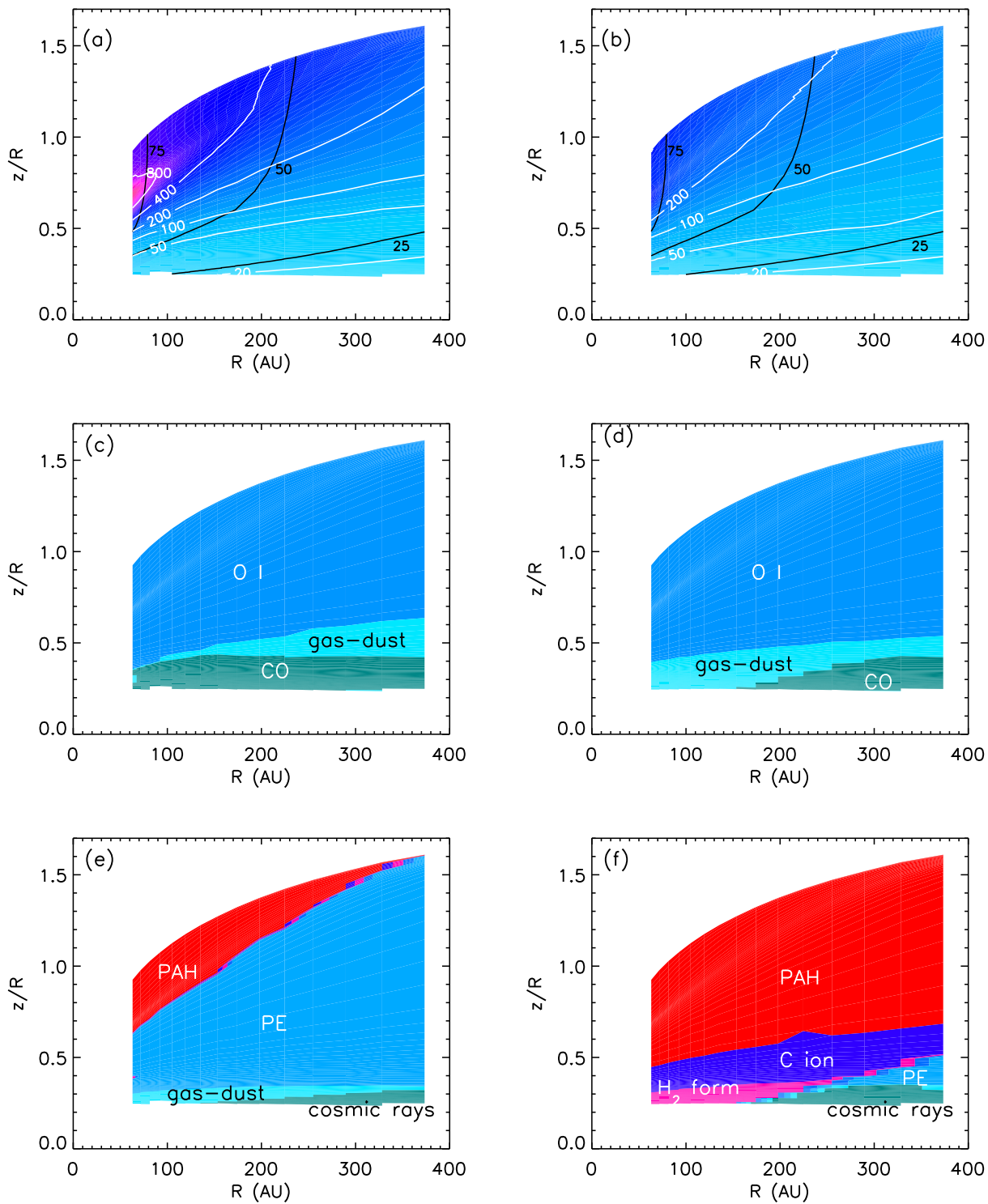}
\caption{The temperature structure of the disk. The colorscale and the white
contours give the gas temperature, the black contours give the dust
temperature for the well-mixed model (a) and the dust settling model
(b). The dominant cooling rates are also shown for the well-mixed (c)
and settled (d) models, as well as the dominant heating rates in (e) and
(f).}
\label{temps1}
\end{figure*}

\begin{figure*}[!tp]
\centering
\includegraphics[width=17cm]{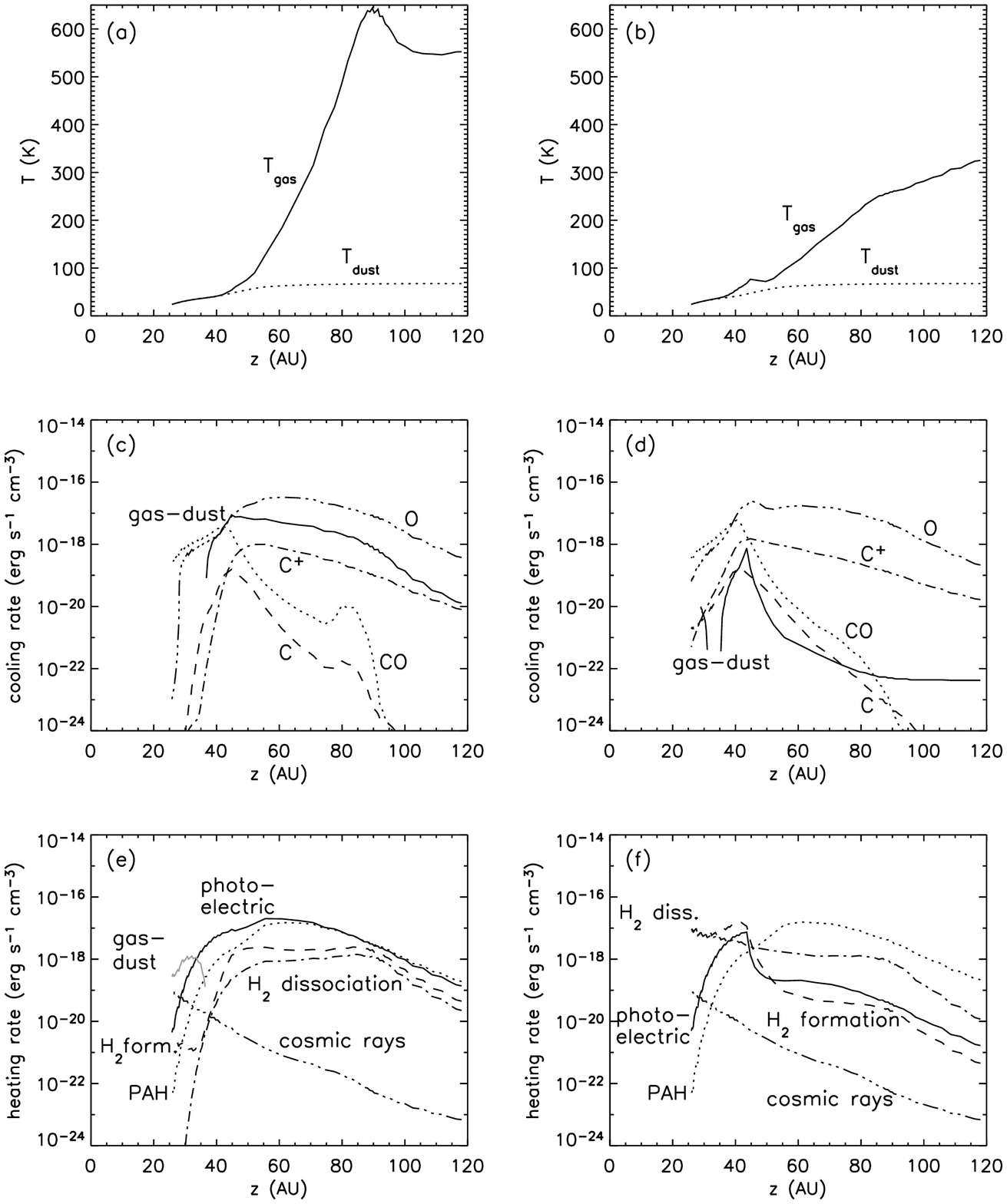}
\caption{The vertical temperature distribution at a radius of 105 AU
for the well-mixed (a) and the settled (b) models. Figures (c) and (d)
give the cooling rates at this radius, where the solid line denotes 
cooling by gas-dust collisions, the dashed line [\ion{C}{i}] cooling, the 
dotted line CO cooling, the dash-dot line [\ion{C}{ii}] cooling and the
dash-triple dot line [\ion{O}{i}] cooling. Figures (e) and (f) give the 
heating rates: the solid line gives the photoelectric heating rates on 
large grains, the dotted line on PAHs; the dashed and dash-dot lines gives the 
heating rates due to the formation and dissociation of ${\rm H_2}$, 
respectively, the dash-triple dot 
line gives the cosmic ray heating rate, and the grey solid line gives the 
heating rate due to gas-grain collisions.}
\label{temps2}
\end{figure*}

It is further assumed that the smallest particles represented by PAHs
remain well-mixed with the gas in all models considered
here. Observations show evidence for PAH emission from disks, at least 
around Herbig Ae stars
\citep[e.g.][]{meeus01}, whereas detailed models indicate that their
settling times are much longer than the ages of the disks
\citep{weiden97}. This means that PAH heating is still at full
strength in the upper layers of the disk while photoelectric heating
on large grains is suppressed in the case of dust settling. The PAHs
are also responsible for approximately half of the absorption of UV
radiation with wavelengths shorter than 1500 \AA \ which dissociates
molecules \citep[e.g.][]{ligree97}. Thus, much of the UV radiation which
affects the chemistry is still absorbed in the upper layers even when
large dust particles are settling. In our models, this is taken into
account by adopting an effective $A_{V}$ in the calculation of the
depth-dependent photodissociation rates. Specifically, the 
gas/dust parameter $x_{\rm gd}$ is defined as the actual value of the 
$m_{\rm gas}/m_{\rm dust}$ divided by the interstellar value. Thus, 
$x_{\rm gd}$ varies between 1 and 100. The visual extinction then 
becomes 
$$A_V=\left(\frac{N_{\rm H}}{1.59\, 10^{21}\,{\rm cm}^{-2}}\right)\times\frac{1}{x_{\rm gd}}$$
The effective extinction used for photorates at UV wavelengths
$$A_{V, {\rm eff}}=\left(\frac{N_{\rm H}}{1.59\, 10^{21}\,{\rm cm}^{-2}}\right)\times\frac{1}{2}\left(\frac{1}{x_{\rm gd}}+1\right)$$
Thus the dissociating radiation is still reduced (albeit at half strength)
in areas where large grains have disappeared completely.

The scaling of the various heating and cooling rates in the settling
model is described in Appendix A. In particular, it should be noted
that although the H$_2$ grain surface formation rate is reduced, ${\rm H_2}$
formation through the reaction ${\rm PAH\!:\!H\,+\,H \to PAH\,+\,H_2}$ is 
included in the chemical network. Since this reaction has a rate comparable 
to the ${\rm H_2}$ formation rate on large grains the abundance of ${\rm H_2}$
is lowered by only a factor of 2 in the upper layers if the settled disk. If
${\rm H_2}$ formation through PAHs is excluded, the ${\rm H\to H_2}$ transition
occurs deeper in the disk, but the effect on the ${\rm C^+\to C\to CO}$ 
transition is small. The overall gas density structure is kept the 
same as in the well-mixed model, as is the dust temperature
distribution. The latter assumption is certainly not valid, but since
gas-dust coupling only plays a role in the lower layers, this does not
affect our results. 

\section{Results}

\subsection{Temperature}

The results of the temperature calculations are presented in Figure
\ref{temps1} for the entire disk, and in Figure \ref{temps2} for a
vertical slice through the disk at 105 AU. It can be seen that the
vertical temperature distribution resembles that of a ``normal'' PDR
\citep[see for example][]{tiehol85}: the gas temperature is much
higher than the dust temperature at the disk surface, and both
temperatures decrease with increasing visual extinction. In
annuli close to the central star the surface temperatures go up to 1000
K.  The precise temperature in these annuli is uncertain and depends
on the adopted molecular parameters.  For example, when the ${\rm
H_2^*}$ vibrational de-excitation rate coefficients given by 
\citet{tiehol85} are used, the rate increases steeply with increasing 
temperature with the cooling rate unable to keep up, resulting in a numerical 
instability in the code. When the rate coefficients given by \citet{lebour99} 
are used, a stable temperature can be found.

\begin{figure}[!htb]
\resizebox{\hsize}{!}{\includegraphics[angle=0]{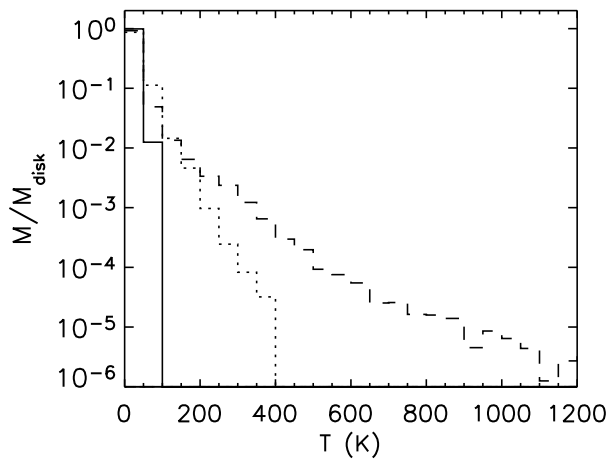}}
\caption{The normalized distribution of the temperature over total disk mass. 
The solid line shows the dust temperature, the dashed line the gas temperature 
of the well-mixed model, and the dotted line the gas temperature of the settled
model.}
\label{temphist}
\end{figure}

The main difference between the temperature structures in disks and 
general molecular clouds is caused by the
increasing density in disks: in standard PDRs the density generally
stays uniform, causing the gas temperature to fall below the dust
temperature at high optical depths. In disks, the increasing density
causes the coupling between gas and dust to dominate the thermal
balance in the deeper layers, so that the gas temperature becomes
equal to the dust temperature.

The normalized distribution of the gas temperature over the disk mass 
between 63 AU and 373 AU of the central star is shown in Figure
\ref{temphist}. The calculations are performed down to a dust
temperature of $\sim 20$ K. It can be seen that even though
the bulk of the gas mass is at the same temperature as the dust, a
considerable fraction ($\sim 10\%$) in the surface layers is at higher 
temperatures,
especially in the case of dust settling. Since most of the molecular
emission arises from this layer, the difference is significant. It can 
also be seen that the settled model contains the largest amount of hot
gas, even though the maximum temperature is higher in the well-mixed
case.

The above figures illustrate that $T_{\rm gas}=T_{\rm dust}$ is
invalid in the upper layers of the disk. This can also be seen in the
individual heating and cooling rates in Figure~\ref{temps2}: gas-dust
collisions dominate the heating/cooling balance only deep within the
disk. In the upper layers the heating due to the photoelectric effect
on large grains and PAHs is so large that cooling of the gas through
collisions with dust grains is not effective anymore, and atomic
oxygen becomes the dominant cooling agent. This shifts the equilibrium 
to higher gas temperatures.

\begin{figure*}
\centering
\includegraphics[width=17cm]{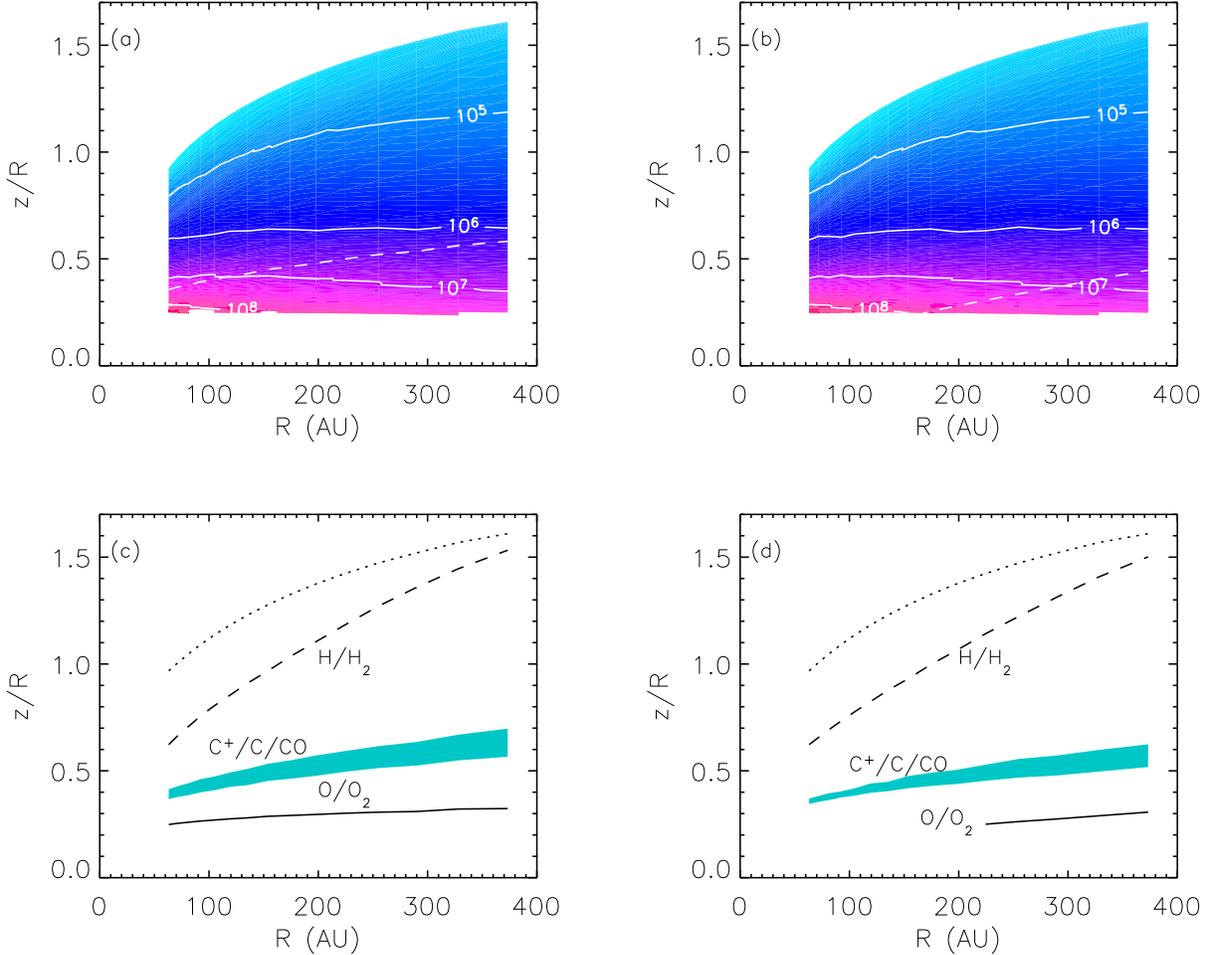}
\caption{Chemical abundances for the well-mixed (left column) and
settled (right column) models. In figures (a) and (b), the density
profile of the disk is shown, as well as the $A_{\rm V}$=1 surface
denoted by the dashed line. In figures (c) and (d), the surface
of the disk is shown as a dotted line, the H/H$_2$ transition as a
dashed line, and the O/O$_2$ transition as a solid line. The
shaded area denotes the C$^+$/C/CO transition.}
\label{2dchem}
\end{figure*}

\begin{figure*}
\centering
\includegraphics[width=17cm]{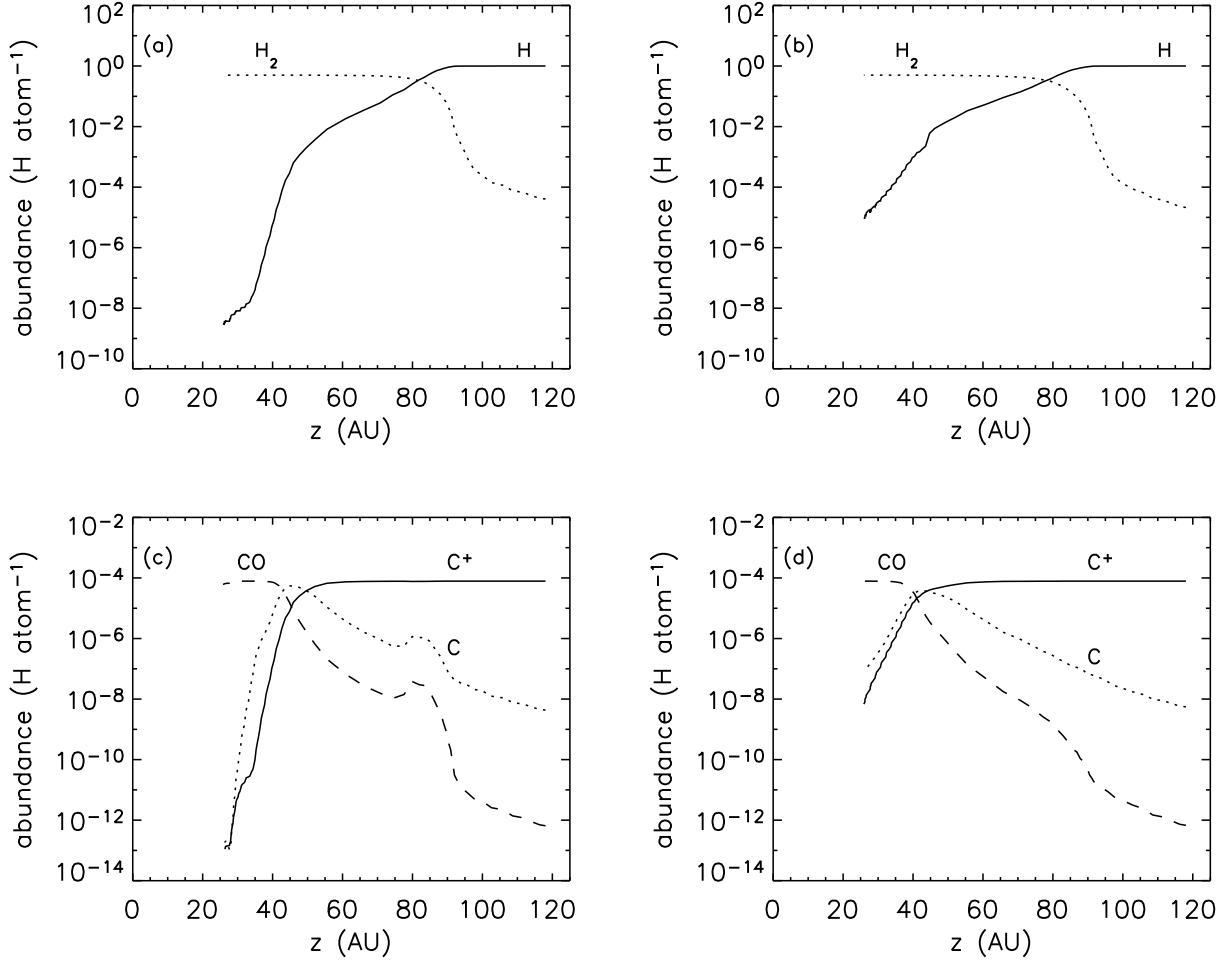}
\caption{The vertical distribution of several chemical species in the
disk at 105 AU for the well-mixed (left column) and settled (right
column) models. Figures (a) and (b) show the abundances of H and ${\rm
H_2}$; figures (c) and (d) of ${\rm C^+}$, C and CO.}
\label{1dchem}
\end{figure*}

In the case of dust settling, the photoelectric effect on large grains
and the ${\rm H_2}$ formation heating rates are suppressed by a factor 
of 100. The PAH heating is still effective, resulting in a temperature
that is $\sim 50\%$ lower than in the well-mixed model. In this model there
is also less absorption of UV radiation in the upper layers, resulting
in higher temperatures deeper in the disk compared to the well-mixed
model. When the abundance of large grains rises, all this UV radiation
is available for photoelectric heating, resulting in a peak in the gas
temperature. This extra UV is rapidly absorbed here, causing a sudden
drop in the temperature as gas-dust collisions become the dominant
process in the thermal balance. If the PAHs are removed from the model,
thus simulating a disk where these small grains have disappeared, the
temperature in the outer regions drops even further, but the overall shape 
of the temperature structure does not change much.

\subsection{Chemistry}

In Figures \ref{2dchem} and \ref{1dchem} the chemical structure of the
disk is presented. The chemistry follows that of an ordinary PDR: a region
consisting mainly of atomic H in the outer layers, with a sharp
transition to the deeper molecular layers. Self-shielding of ${\rm
H_2}$ quickly reduces the photodissociation rate, so the deeper layers
consist mainly of molecular hydrogen. The principle
carbon-bearing species follow a similar trend, consisting mainly of
${\rm C^+}$ in the upper layers, with a transition to C, and later CO,
at larger optical depths. In the deeper layers CO is also 
protected from dissociation by self-shielding and by shielding by 
${\rm H_2}$, as several dissociative wavelengths overlap for these 
molecules. Because CO is
much less abundant than ${\rm H_2}$, absorption by dust also plays an
important role in decreasing its dissociation rate. This causes the
carbon in the disk to be mostly ionic in the upper layers, and mostly
molecular (in the form of CO) deeper in the disk. Compared with the 
models of \citet{aikawa02} and \citet{zade03}, 
our C $\to$ CO transition occurs somewhat
deeper into the disk, at $z\sim$40 AU rather than 60 AU for the
$R$=105 AU annulus. This is due to the geometry in the 1+1-D model, which 
allows dissociating radiation to penetrate more deeply into the disk, 
as well as our different treatment of CO photodissociation.

In the case of dust settling the ${\rm C^+}$/C/CO transition occurs
 slightly deeper in the disk. This
is explained by the higher photodissociation rates
deeper in the disk due to the decreased absorption of UV radiation by
large dust grains. The PAHs in the outer regions still absorb a significant 
fraction of the UV, so the effect is not very large. Since the H/${\rm H_2}$ 
transition is determined by ${\rm H_2}$ self shielding and not dust absorption,
the position of this transition does not change if the dust settles.

The chemistry has also been calculated for a model where 
$T_{\rm gas}=T_{\rm dust}$. It is found that the different temperature 
has little effect on the abundances of those species important in the thermal
balance. The chemistry in the surface layers of disks is mostly 
driven by photo-reactions and ion-molecule reactions, both of which are 
largely independent of temperature.

\subsection{Line intensities}

\begin{figure*}[!tp]
\centering
\includegraphics[width=17cm]{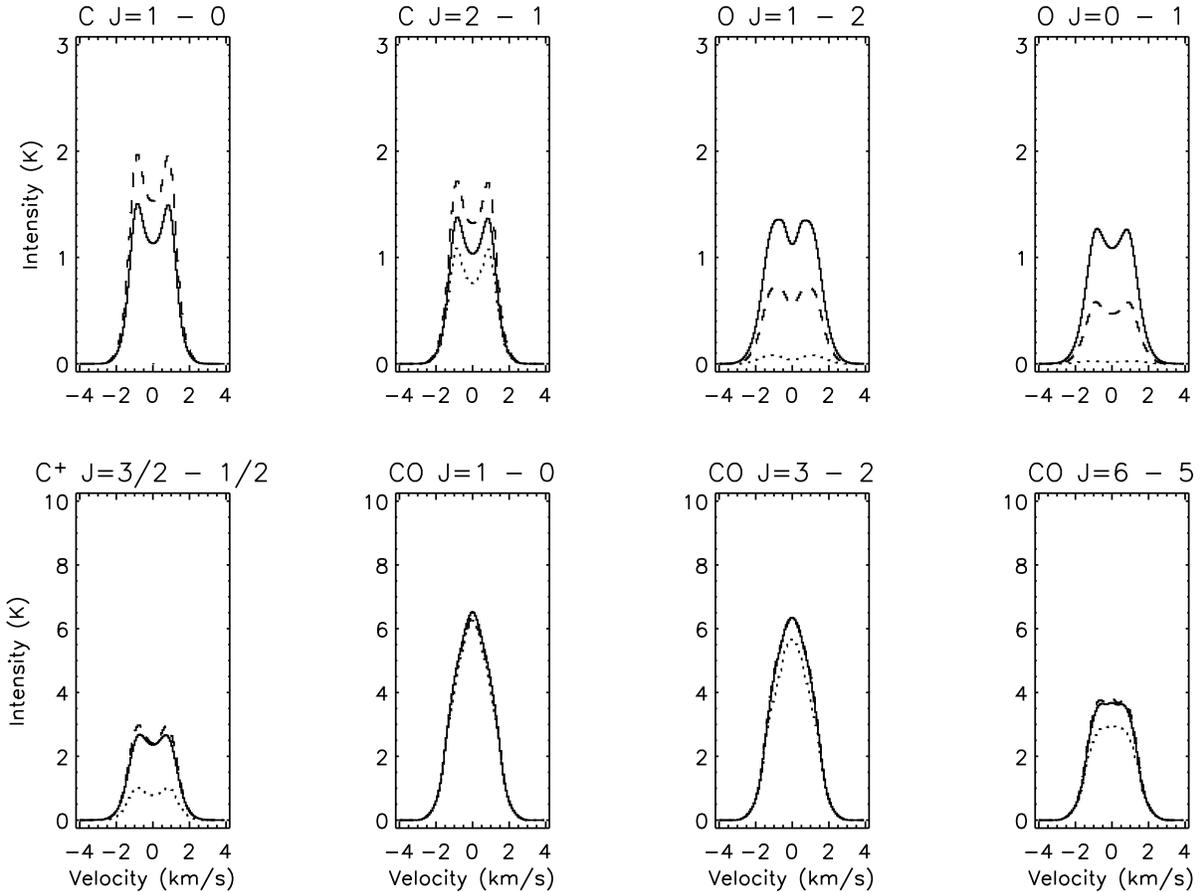}
\caption{The emission lines of C, O, ${\rm C^+}$ and CO for three scenarios: $T_{\rm gas}=T_{\rm dust}$ (dotted line), and $T_{\rm gas}$ calculated explicitly in the well-mixed (solid line) and settled (dashed line) model. Units are in Kelvins, assuming that the disk fills the beam precisely; this corresponds to a beam size of $5\arcsec$ when the disk is at 150 pc.}
\label{specs}
\end{figure*}

The main effect of the high gas temperatures is on the intensities and
shapes of emission lines arising from the warm surface layers. It is
expected that the intensities of lines tracing high temperatures (such
as the [\ion{O}{i}] fine-structure lines, the ${\rm H_2}$ rotational lines,
and the higher CO rotational lines) will increase due to the higher
temperatures, and the larger overall amount of warm gas. 

The model line profiles are created using the 2-D Monte Carlo code by 
\cite{hogtak00} assuming a Keplerian velocity field
and an inclination of $60^\circ$. The results for the 
[\ion{C}{i}], [\ion{O}{i}] and [\ion{C}{ii}] fine-structure lines and 
the rotational lines of CO are shown in 
Figure \ref{specs}. It can be seen that the high gas temperatures have a 
significant impact on the intensities, and particularly on the [\ion{O}{i}] 
 and [\ion{C}{ii}] fine-structure lines. The difference in the locations in 
the disk where these lines are predominantly excited is reflected in the
different shapes of the lines. The lines tracing hot gas all display a 
double-peaked structure, due to their excitation in the innermost parts
of the disk. It can also be seen that the 
higher gas temperature has little effect on the lower rotational lines of 
CO and the lowest [\ion{C}{i}] fine-structure line. These lines trace the 
cold gas, which is present in the deeper layers of the disk in all models.

\section{Discussion}
\subsection{Limitations of the 1+1-D model}

While circumstellar disks are inherently 2-dimensional
structures, a complete 2-D treatment of the radiative transfer, chemistry 
and thermal balance is at present too time-consuming to be practical.
The 1+1-D model circumvents this problem by drastically simplifying the 
geometry to a series of 1-D structures. The main disadvantage of the 1+1-D 
model is its poor treatment of radiative transfer: in principle, only 
transfer in the vertical direction is included. In a pure 1+1-D geometry
the stellar radiation is forced to change direction when it hits the disk's 
surface and thus affects the chemistry calculation. This problem does not 
apply to the photoeletric heating rates of PAHs and large grains since they
were  calculated using a UV field calculated with the 2-D Monte Carlo code by 
\citet{zade03}.

The 1+1-D geometry also affects the escape probabilities of the
radiative cooling lines, which have to travel vertically instead of
escaping via the shortest route to the disk's surface. Thus the escape
probabilities calculated by the 1+1-D model are generally too low. 

The 
overall effect of the 1+1-D geometry on the disk temperature is
subtle: the very upper layers of the disk are optically thin
regardless of the adopted geometry, so the temperature 
and chemistry results are expected to be correct there. In the dense 
regions near the midplane, the dust opacities are so large that there will 
be little effect of the geometry either. Also, due to the strong thermal 
coupling between gas and dust, the gas temperature is equal to the dust 
temperature there. The greatest uncertainties are in the 
chemistry, in the intermediate regions between the surface
and the midplane. Since the thermal balance is tied to the chemistry, it is
expected that the largest uncertainties in the gas temperature occur also in 
this region. It should be noted that many secondary effects play a
role here, including the precise formulation of the cooling rates, 
treatment of chemistry, self-shielding,
etc. The current models should be adequate to capture the main
characteristics and magnitude of all of these effects.

\subsection{Effects of gas-dust coupling}

\begin{figure}
\resizebox{\hsize}{!}{\includegraphics[angle=0]{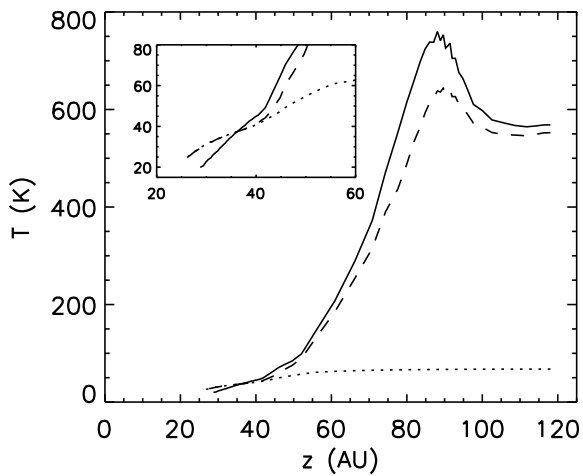}}
\caption{Vertical distribution of gas and dust temperatures in the
disk at a radius of 105 AU. The solid line gives the gas temperature
when gas-dust collisions are ignored, the dashed line gives the gas
temperature when these collisions are included in the thermal
balance. The dotted line gives the dust temperature. The inset shows
the region around $T_{\rm gas}=20\ {\rm K}$.}
\label{gdtemp}
\end{figure}

It can be seen in Figure \ref{gdtemp} that thermal coupling between
gas and dust has a small effect on the gas temperature in the upper
layers of the disk: the gas temperature is so high  and density comparitively 
low that gas-dust collisions are an ineffective cooling mechanism.
Deeper in the disk the coupling becomes stronger and eventually comes to 
dominate the thermal balance.
The overall effect on the resulting temperature is small because
gas and dust already have similar temperatures even when the coupling
is ignored. 

\subsection{Effects of the radiation field}
The radiation field used in this work is assumed to have the same spectral
shape as the \citet{draine78} field with integrated intensity matched to 
the observed radiation field of TW Hya \citep{costa00}. \cite{bergin03}
have shown that this treatment may overestimate the amount of radiation 
in the 912-1100 \AA \ regime where ${\rm H_2}$ and CO are dissociated,
since a significant fraction of the UV flux is in emission lines (particularly 
Ly~$\alpha$). Thus the dissociation rates of ${\rm H_2}$ and CO, as well as 
the C ionization rate may be too high in this work. Other molecules such as 
H$_2$O and HCN, however, have significant cross sections at Ly$\alpha$.

If the radiation field of \citet{bergin03} is used, the chemistry of the 
relevant cooling species will become
more similar to that found using spectrum C (a 4000 K blackbody) in
\citet{zade03}. Because of the different chemistry, it
is also expected that the temperature profile will change: CO will
become the dominant carbon-bearing species at larger heights, where it
can cool the gas more efficiently. It is therefore expected that the
temperature in the intermediate layers will drop with respect to the
results presented here. The heating rates are not sensitive to changes 
in the chemistry, so they will not change much.

\section{Conclusions}

The main conclusions of our calculations are:
\begin{itemize}
\item The gas temperature is much higher than the
dust temperature in the optically thin part of the disk, while the two
temperatures are the same in the part of the disk which is optically
thick to UV radiation and where most of the disk's material resides. 

\item The temperatures in the optically thin part have a noticeable 
effect on the excitation of higher-lying emission lines. 
The higher temperatures should also affect the disk's structure: 
since the structure is
calculated assuming hydrostatic equilibrium, the high temperatures
in the surface regions will blow-up the disk, which will achieve hydrostatic
equilibrium at a greater height. This in turn will lead to a more
efficient outgassing of the disk. An initial discussion of this effect is
given in \citet{kamdul04}.

\item The chemistry in the disk does not change much when the gas
temperature is increased in the explicit calculation.  

\item Dust settling is found to have a great impact on the gas
temperature in the disk. The gas temperature in disks
where settling is taking place is found to be lower, and decreases
less steeply than in well mixed disks. Due to the lower temperatures, the
intensities of higher-lying lines are lower than in well mixed disks.

\item Dust settling does not greatly affect the disk's chemistry. This
is because PAHs are assumed to stay well-mixed with the gas 
when large grains are settling. As a result much of UV radiation important
for chemistry is still absorbed in the surface layers of the disk in the
settled model. The ${\rm C^+}$/C/CO and the O/${\rm O_2}$ transitions thus
occur only slightly lower in the disk.

\item Dust settling increases the intensity of atomic and molecular  lines, 
and has a subtle influence on the line shapes due to the smaller amount 
of hot gas, especially at small radii.
\end{itemize}

\begin{acknowledgements}

The authors are grateful to Inga Kamp for many discussions on the
thermal balance, and for jointly carrying out a detailed comparison of
codes.  They thank Michiel Hogerheijde and Floris van der Tak for the
use of their 2-D radiative transfer code.  This work was supported by
a Spinoza grant from the Netherlands Organization of Scientific
Research (NWO) and by the European Community's Human Potential Programme
under contract HPRN-CT-2002-00308, PLANETS.

\end{acknowledgements}

\bibliographystyle{aa}
\bibliography{paper1.bib}

\appendix

\section{Heating processes}
In the following, the various processes included in the calculation of
the total heating rate are discussed. Also, their scaling with $m_{\rm
gas}$/$m_{\rm dust}$ is indicated.

{\it Photoelectric heating:} the energetic electrons released by
absorption of UV photons by dust grains contribute significantly to the total
heating rate in the surface of the disk. For the heating rate due to
the photoelectric effect the formula by \cite{tiehol85} is used:
$$
\begin{array}{c}
\Gamma_{\rm PE} = 2.7\times 10^{-25}\,\delta_{uv}\,\delta_{d}\,n_{\rm H}\, Y\, I_{\rm UV}\\
\\
\times\,[(1-x)^2/x\,+\,x_k(x^2-1)/x^2]\ {\rm erg\,cm^{-3}\,s^{-1}}
\end{array}
$$
with $\delta_{uv}=2.2$, $\delta_d=1.25$ and $Y=0.1$. The strength of the radiation field $I_{\rm UV}$ is calculated with the 2-D Monte Carlo code by \cite{zade03}. $x$ is the grain charge 
parameter, found by solving the equation
$$x^3\,+\,(x_k\,-\,x_d\,+\,\gamma)x^2\,-\,\gamma\,=\,0$$ 
where $x=E_0/E_H$, $x_k=kT/E_H$, $x_d=E_d/E_H$ and $\gamma=2.9\times 10^{-4}
Y \sqrt{T} I_{\rm UV} n_e^{-1}$; $E_H$ is the
ionization potential of hydrogen, $E_d$ is the ionization potential
of a neutral dust grain (6 eV is assumed here) and $E_0$ is the grain
potential including grain charge effects. This rate is scaled with
$m_{\rm dust}/m_{\rm gas}$ to account for the varying gas to dust
ratio in models where dust settling was included.\\

{\it PAH heating:} analogous to photoelectric heating, the
photoelectrons from PAH ionization also contribute to the heating of
the gas. The heating rate described in \citet{baktie94} is used,
using PAHs containing $N_C$ carbon atoms $(30<N_{\rm C}<500$). 
The radiation field used to obtain this heating rate is calculated 
with the 2-D Monte Carlo code by \cite{zade03}.\\

{\it C photoionization:} it is assumed that each electron released by
the photoionization of neutral C delivers approximately 1 eV to the
gas. The heating rate then becomes
$$\Gamma_{\rm C\ ion} = 1.7\times 10^{-12}\, R_{\rm C\ ion}\,n({\rm C})\, {\rm erg\,cm^{-3}\,s^{-1}}$$
where $R_{\rm C\ ion}$ is the C ionization rate.\\

{\it Cosmic rays}: another source of energetic electrons is the
ionization of H and $\mathrm{H_2}$ by cosmic rays. It is assumed that
about 8 eV of the electron's energy is used to heat the gas in the
case of $\mathrm{H_2}$, and 3.5 eV in the case of H. The ${\rm H_2}$
contribution is scaled to include the ionization of He. A cosmic ray
ionization rate of $\zeta_{\rm CR}=5\times 10^{-17}\,{\rm s^{-1}}$ is
used. The heating rate then becomes
$$
\begin{array}{c}
\Gamma_{\rm CR} = [2.5\times 10^{-11}\, n({\rm H_2})\, + \, 5.5\times 10^{-12}\,n({\rm H})]\\
\\
\times\,\zeta_{\rm CR}\, {\rm erg\,cm^{-3}\,s^{-1}}
\end{array}
$$

{\it $\mathrm{H_2}$ formation:} of the 4.48 eV liberated by formation
of $\mathrm{H_2}$ on dust grains, it is assumed that 1.5 eV is
returned to the gas. The corresponding heating rate is:
$$\Gamma_{\rm H_2\ form} = 2.4\times 10^{-12}\,R_{\rm H_2\ form}\,
n_{\rm H}\, {\rm erg\,cm^{-3}\,s^{-1}}$$ where $R_{\rm H_2\
form}=3\times 10^{-18}\,\sqrt{T}\,n_{\rm H}\,n({\rm H})/(1 + T/T_{\rm
stick})$ is the ${\rm H_2}$ formation rate, with $T_{\rm stick}$=400
K. This rate is scaled with $m_{\rm dust}/m_{\rm gas}$ in models where
dust settling is included.\\

{\it $\mathrm{H_2}$ dissociation:} when $\mathrm{H_2}$ is excited into
the Lyman and Werner bands, there is about 10\% chance that it will
decay into the vibrational continuum, thus dissociating the
molecule. Each of the H atoms created this way carries approximately
0.4 eV. This gives:
$$\Gamma_{\rm H_2\ diss} =  6.4\times10^{-13}\, n({\rm H_2})\, R_{\rm H_2\ diss}\, {\rm erg\,cm^{-3}\,s^{-1}}$$
where $R_{\rm H_{2\ diss}}$ is the ${\rm H_2}$ photodissociation rate.\\

{\it $\mathrm{H_2}$ pumping:} electronically excited
$\mathrm{H_2}$ that does not disscociate decays into bound
vibrationally excited states of the electronic ground state. It is
assumed that 2.6 eV is returned to the gas by collisional
de-excitations, resulting in:
$$\Gamma_{\rm H_2 pump}= 4.2\times 10^{-12}\, [n({\rm H})\,\gamma_{\rm H}\,+\,n({\rm H_2})\,\gamma_{\rm H_2}]$$
$$\times\,n({\rm H_2^\ast})\,{\rm erg\,cm^{-3}\,s^{-1}}$$ where
$\gamma_{\rm H}$ and $\gamma_{\rm H_2}$ are the deexcitation rate coefficients
through collisions by H and ${\rm H_2}$, respectively. These rates are
taken from \citet{lebour99}. H$_2^*$ is calculated explicitly in our models
in the UV excitation of H$_2$.\\

{\it Chemical heating:} in principle every exothermic reaction
contributes to the heating of the gas. The reactions considered here
(and the energies released) are the dissociative recombination of 
$\mathrm{H_3^+}$ (in which 9.23 eV is realeased by dissociating to 
H+${\rm H_2}$ and 4.76 eV by dissociating to H+H+H), $\mathrm{HCO^+}$ 
(7.51 eV) and $\mathrm{H_3O^+}$ (6.27 eV), and the destruction by
$\mathrm{He^+}$ ions of $\mathrm{H_2}$ (6.51 eV) and CO (2.22 eV).

{\it Gas-dust collisions:} collisions between gas and dust will heat
the gas when the dust temperature is higher than the gas temperature,
and will act as a cooling term when the dust temperature is lower than
the gas temperature. The formula by \cite{burhol83} is used, assuming
an average of $\left<\sigma_{\mathrm gr}n_{\mathrm gr}\right>=1.5\,
n_{\rm H}\,\mathrm{cm^{-1}}$ and an accomodation coefficient
$\bar{\alpha}_T = 0.3$. The resulting rate is scaled with the local
value of $m_{\rm dust}/m_{\rm gas}$ in models where dust settling was
included:
$$\Gamma_{\rm gd} = 1.8\times 10^{-33}\,n_{\rm H}^2\, \sqrt{T}\, (T_{\rm dust}-T_{\rm gas})\, {\rm erg\,cm^{-3}\,s^{-1}}$$

\section{Cooling processes}

The gas is cooled through the fine-structure lines of $\mathrm{C^+}$,
C and O, and the rotaional lines of CO. For each species the level
population was calculated assuming statistical equilibrium:
$$\frac{dn_i}{dt}=\sum_{j\neq i}n_j\, R_{j\to i}\ -\ n_i\sum_{j\neq
i}R_{i\to j}\,=\,0$$ where $n_i$ is the density of a given species in
energy state $i$, and $R_{i\to j}$ is the total rate of level $i$ to
level $j$. For all species, collisional excitations and de-excitations
are taken into account, as well as spontaneous emission and
absorption and stimulated emission through interaction with the cosmic
microwave background (CMB) and the dust infrared background. For ${\rm
C^+}$ and O, UV pumping of the fine-structure lines is also
included. Once the populations are known, the cooling rate can be found
using the formula by \cite{tiehol85}:

$$\Lambda_X(\nu_{ij})=n_i\,A_{ij}\,h\nu_{ij}\,\beta_{\rm
esc}(\tau_{ij})\frac{S(\nu_{ij})-P(\nu_{ij})}{S(\nu_{ij})}$$ In this
formula, $n_i$ is the density of species $X$ in energy state $i$,
$A_{ij}$ is the Einstein A coefficient for the transition $i\to j$,
$\nu_{ij}$ is the corresponding frequency, and $\beta_{\rm
esc}(\tau_{ij})$ is the escape probability at optical depth
$\tau_{ij}$. $S(\nu_{ij})$ is the local source function:

$$S(\nu_{ij})=\frac{2h\nu_{ij}^3}{c^2}\left(\frac{g_in_j}{g_jn_i}-1\right)^{-1}$$
where $g_i$ is the statistical weight of level $i$. $P(\nu_{ij})$ is
the intensity of the background radiation:

$$P(\nu_{ij})=B_{\nu_{ij}}(T_{\rm CMB}) + \tau_{\rm
dust}B_{\nu_{ij}}(T_0)$$ where $T_{\rm CMB}$ is the temperature of the
cosmic background (2.726 K) and $T_0$ the characteristic temperature
of the dust at the surface of the disk. In the calculation of  the collision 
rates of all cooling species considered here, the ortho/para ratio for 
${\rm H_2}$ was assumed to be in  thermal equilibrium with the local gas 
temperature.

{\it O cooling:} O contributes to the cooling of the gas in the
surface layers of the disk through its fine-structure lines at ${\rm
63.2\,\mu m}$ and ${\rm 145.6\,\mu m}$. The Einstein A coefficients
for these lines are $8.87\times 10^{-5}\, {\rm s^{-1}}$ and 
$1.77\times 10^{-5}\, {\rm s^{-1}}$, respectively. Collisions with electrons
\citep{haynus84}, H \citep{laurou77a} and ${\rm H_2}$ \citep{jaquet92}
were included. O can also act as a heating agent if infrared pumping is
followed by collisional de-excitation. If that happens the cooling rate
becomes negative and is treated as a heating rate by the code. This
phenomenon did not occur in the calculations presented in this paper.

{\it $\mathrm{C^+}$ cooling:} $\mathrm{C^+}$ contributes to the
cooling of the gas through its fine-structure line at $158\,{\rm \mu
m}$. An Einstein A coefficient of $2.29\times 10^{-6}\, {\rm s^{-1}}$
is used. Collisions with electrons \citep{haynus84}, H
\citep{laurou77b} and ${\rm H_2}$ \citep{flolau77} are included in
calculating the level populations.

{\it C cooling:} C cools the gas through its fine-structure lines at ${\rm
370\, \mu m}$ and ${\rm 609\, \mu m}$. The Einstein A coefficients
used are $2.68\times 10^{-7}\, {\rm s^{-1}}$ and $7.93\times
10^{-8}\, {\rm s^{-1}}$, respectively. Collisions with H
\citep{laurou77a} and ${\rm H_2}$ \citep{schrod91} are included.

{\it CO cooling:} CO is the main coolant of the dense, shielded regions deep
in the disk through its rotational lines. The 20 lowest rotational
transitions of CO are included in the model. The Einstein A
coefficients used in this work are identical to those adopted by
\citet{kamzad01}. Collisions with H \citep{wabevi96} and ${\rm H_2}$
\citep{flower01} are taken into account to calculate the
populations. 

\end{document}